\documentclass[11pt,preprint,preprintnumbers,amsmath,amssymb,nofootinbib,aps]{revtex4}
\usepackage{graphicx}
\usepackage{hyperref}
\usepackage{amsmath}
\usepackage{amssymb}
\usepackage{amsfonts}
\usepackage[justification=justified]{subcaption}
\usepackage[justification=justified]{caption}
\newcommand{\be}{\begin{equation}}
\newcommand{\ee}{\end{equation}}
\newcommand{\bi}{\begin{itemize}}
\newcommand{\ei}{\end{itemize}}
\newcommand{\bea}{\begin{eqnarray}}
\newcommand{\eea}{\end{eqnarray}}

\begin{document}

\title{Cytoskeletal turnover and Myosin contractility drive cell autonomous oscillations in a model of Drosophila Dorsal Closure.}

\author{Pedro F. Machado}
\email{pf288@gen.cam.ac.uk}
\affiliation{Department of Genetics, University of Cambridge, Downing Street, Cambridge CB2 3EH, UK.}

\author{Guy B. Blanchard}
\email{gb288@cam.ac.uk}
\affiliation{Department of Physiology, Development and Neuroscience, University of Cambridge, Downing Street, Cambridge CB2 3DY, UK.}

\author{Julia Duque}
\email{jduque@cbm.uam.es}
\affiliation{Centro de Biolog\' {i}a Molecular \it{Severo Ochoa}, CSIC, C/ Nicol\'{a}s Cabrera 1, 28049 Madrid, Spain}

\author{Nicole Gorfinkiel}
\email{ngorfinkiel@cbm.uam.es}
\affiliation{Centro de Biolog\' {i}a Molecular \it{Severo Ochoa}, CSIC, C/ Nicol\'{a}s Cabrera 1, 28049 Madrid, Spain}

\begin{abstract}

Oscillatory behaviour in force-generating systems is a pervasive phenomenon in cell biology. In this work, we investigate how oscillations in the actomyosin cytoskeleton drive cell shape changes during the process of Dorsal Closure, a morphogenetic event in \textit{Drosophila} embryo development whereby epidermal continuity is generated through the pulsatile apical area reduction of cells constituting the amnioserosa (AS) tissue. We present a theoretical model of AS cell dynamics by which the oscillatory behaviour arises due to a coupling between active myosin-driven forces, actin turnover and cell deformation. Oscillations in our model are cell-autonomous and are modulated by neighbour coupling, and our model accurately reproduces the oscillatory dynamics of AS cells and their amplitude and frequency evolution. A key prediction arising from our model is that the rate of actin turnover and Myosin contractile force must increase during DC in order to reproduce the decrease in amplitude and period of cell area oscillations observed in vivo. This prediction opens up new ways to think about the molecular underpinnings of AS cell oscillations and their link to net tissue contraction and suggests the form of future experimental measurements.


\end{abstract}

\maketitle

\section{Introduction}
\label{intro}

The development of organs and tissues in living systems requires the integration of chemical and mechanical processes occurring at subcellular, cellular, and multicellular level \cite{Blanchard&Adams2011}. The actomyosin cytoskeleton plays a central role in morphogenesis through the control of cell shape, the generation of forces and the modulation of material properties \cite{Levayer&Lecuit2013,Salbreux2012}. While much progress has been made in vitro and at the level of single cells to understand how the cytoskeleton impinges onto cell mechanical properties and cellular behavior \cite{Fletcher&Mullins2010}, understanding how the dynamic activity of the cytoskeleton is integrated at the cell and multicellular level remains an open challenge. In this work, we present a phenomenological model of epithelial morphogenesis that captures the effective dynamics of the cytoskeleton, using \textit{Drosophila} Dorsal Closure as our model system.

Dorsal Closure (DC) is a tissue remodelling process whereby a gap at the dorsal side of the \textit{Drosophila} embryo is closed to generate epidermal continuity \cite{Gorfinkiel2011}. This gap is covered by an extraembryonic squamous epithelium called the amnioserosa (AS). Laser ablation and genetic perturbation experiments have identified the AS as providing one of the main contributing forces to DC through the pulsatile reduction in apical area of its constituent cells \cite{Hutson2003,Blanchard2009,David2010,Solon2009}.  At the tissue level, DC progresses along three distinct stages, early stage, with cells exhibiting apical cell area oscillations but no net area reduction, slow stage, in which tissue area contraction begins, and fast stage, characterised by an increase in the tissue contraction rate and coinciding with the onset of zippering \cite{Gorfinkiel2009,Blanchard2009}. At the cellular level, this apical area reduction can be decomposed along two timescales, short timescale area oscillations with a period of $2-4$ minutes and a long timescale net area contraction occurring over the span of one hour. Live imaging shows myosin is the driving force behind apical cell area oscillations and effective contraction, with actomyosin foci periodically accumulating and dispersing along the apical cortex \cite{Blanchard2009,David2010}. These actomyosin oscillations anticorrelate with apical area, with peaks in myosin density preceding the troughs in cell area. Lastly, oscillatory behaviour is patterned in time, with the period and amplitude of area oscillations decreasing as DC progresses \cite{Blanchard2009}.

Thus, DC offers an excellent in vivo system in which to study how the integration of cytoskeletal activity, cellular dynamics and tissue mechanics generates the macroscopic behavior of active matter. Previous experimental and theoretical studies have raised several questions, among which the most notable are the origin of cytoskeletal and cell area oscillations and the evolution of this oscillatory dynamics. In this work, we searched to address these questions via a cell-based model of DC. Recently, a cell-based model has been suggested whereby oscillations are driven by a putative feedback between myosin and a signalling molecule \cite{Wang2012}, stressing a chemical mechanism for oscillations. In a different context however, a mechano-chemical feedback has been suggested to underlie oscillations in cell volume during cytokinesis \cite{Sedzinski2013}.  

Here, we propose an extension of the single-cell model of \cite{Sedzinski2013} to the context of epithelial tissues, whereby oscillations arise due to a coupling between actomyosin turnover, cell deformation and neighbour effects. We show that this coarse-grained description quantitatively reproduces the observed behaviour of AS cells in terms of the evolution of amplitude and period of cell area oscillations. In this model, the oscillatory dynamics is controlled by four parameters, the viscoelastic relaxation timescale, the actomyosin turnover rate, the ratio of active to passive force and the ratio of contractile to passive force. The time evolution of oscillation suggests an evolution in these parameters and a tight control of turnover and active force generation. By performing live imaging of cell membrane and actin reporters and quantifying cell shape change, we find that such an evolution of actomyosin turnover rates could take place during DC. 

\section{Model formulation}
\label{model}

\begin{figure}
\centering
\includegraphics[width=0.8\linewidth]{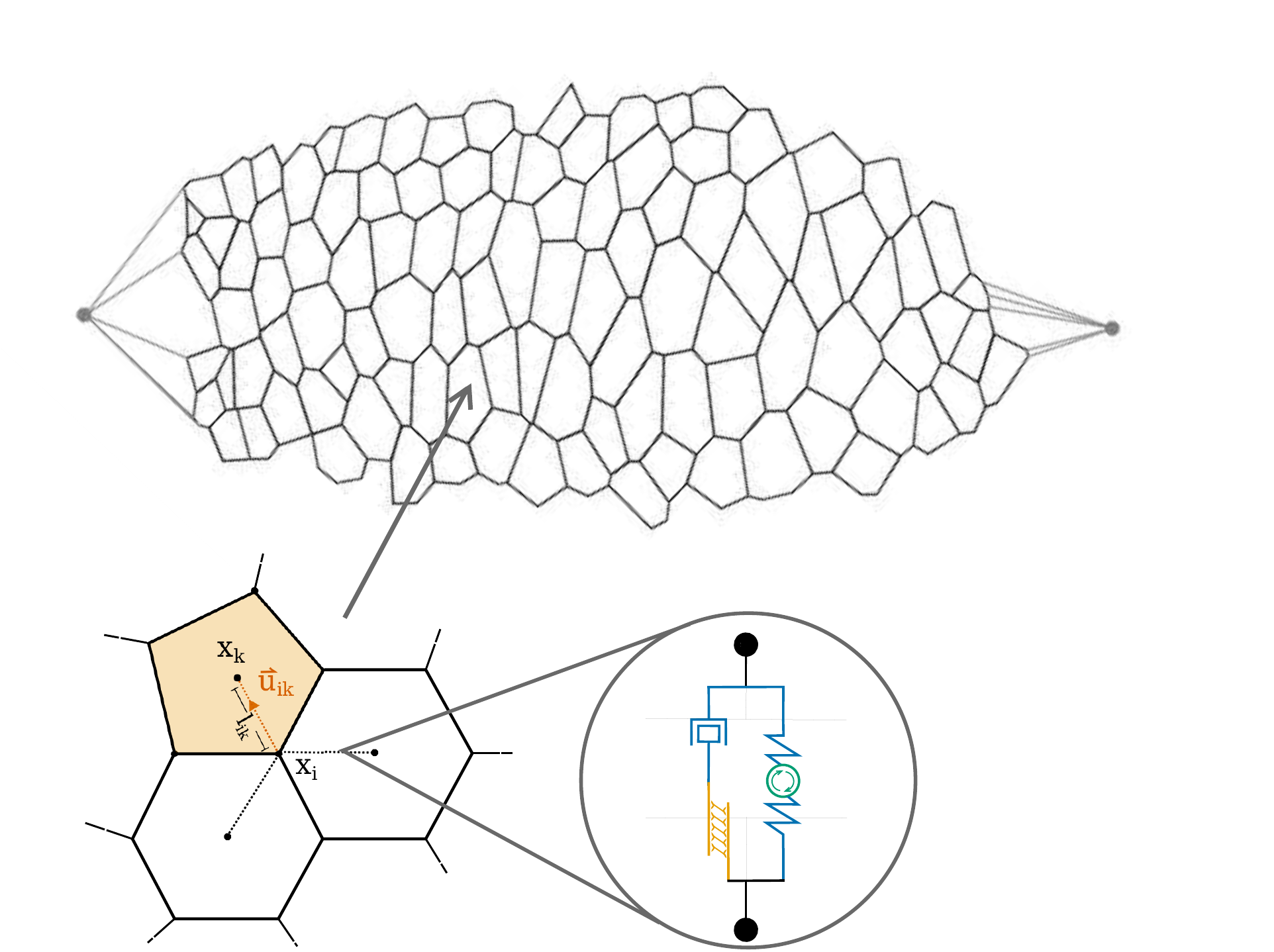}
\caption{\label{fig:forces} {\footnotesize Tissue segmentation and AS model. Top: Early DC segmentation; grey circles correspond to fixed canthi, to which anteriormost and posteriormost boundary vertices are tethered (grey lines). Bottom: cell-level forces in our model. Forces at vertex $x_i$ are resolved along the vertex to centroid (spoke) vectors $u_{ik}$ for every cell $k$ to which the vertex belongs (left).  For each vertex-centroid spoke, net force is the result of passive viscoelastic forces dependent on the spoke length $l_{ik}$ (right inset, blue), active actomyosin-driven forces dependent on actomyosin density $a_k$ (right inset, orange) and a net contractile force (right inset, green). The effect of the net contractile force is to gradually reduce the spoke rest length.}}
\end{figure}

We propose a cell-based model for the AS coupling cell mechanical deformation to actomyosin dynamics. As the AS is a squamous epithelium and apical dynamics has been shown to dominate DC, we adopt a 2D representation of AS cells. We approximate cells by polygons and encode cell mechanics via vertex displacements which occur due to passive viscoelastic forces, active myosin-driven forces and long-timescale contractile forces. Since medial actomyosin has been identified as the key active force element in AS cell deformation and AS cell membranes exhibit wiggly borders up to the onset of the fast phase \cite{Blanchard2009}, suggesting low interface tension, we take these forces to be resolved along vertex to cell centroid vectors only (see Fig.~\ref{fig:forces}, inset). Given that AS cell area oscillations occur around a single frequency order of magnitude and under low strain ($\sim 10\%$) \cite{Blanchard2009,Jayasinghe2013}, we take the passive force to be that of a linear viscoelastic solid with a single relaxation timescale. Lastly, we assume that a threshold amount of myosin concentration is required to generate deformations at the cell level \cite{Alvarado2013} and that the active force component is proportional to a cell's myosin concentration. The total force acting on a cell vertex $i$ is thus given by 
\be\label{eq:motion}
\eta \frac{d\vec{x}_i}{dt} = \sum_{\rm k} \vec{u}_{ik}\left[ \kappa \left(l_{ik}-l_{ik}^{0}\right)+\alpha_m\,f_m\left(\tfrac{M_k}{M^0_k}\right)\right] \,,
\ee
where $\kappa$ is the elastic modulus, $\eta$ is a friction factor, $\alpha_m$ is the tensile force per unit myosin, $l_{ik}$  and $l_{ik}^0$ are the current length and rest length of the spoke connecting the vertex-centroid pair $i$,$k$, respectively, and $\vec{u}_{ik}$ is the unit vector directed towards the centroid $k$. The term $\,f_m\left(\tfrac{M_k}{M^0_k}\right)$ represents the active contractile force and is given by the sigmoidal function
\be\label{eq:ratchet}
f_m\left(\tfrac{M_k}{M_0}\right) = \frac{M_k/M^0_k}{1+e^{-2 n (M_k/M^0_k -1)}}-\frac{1}{2}\,,
\ee
where $M_k$ is the myosin concentration of cell $k$, $M^0_k$ is a threshold concentration and $n$ is an integer setting the slope of the sigmoid, which we set to $n > 4$. Given the difference in timescales between net area reduction and area oscillations in AS cells, we implement the net contractile force via a gradual reduction of the spoke rest length with
\be
\frac{dl_{ik}^0}{dt} = -\nu(t) \,,
\ee
where $\nu(t)$ is a piecewise constant contraction rate. This implementation is similar to the `internal ratchet' of \cite{Wang2012}, which was shown to more accurately reproduce DC dynamics than the external, actin-cable driven ratchet proposed in \cite{Solon2009}. 

The equations of motion of the model are coupled to a set of equations describing the intracellular regulation of actomyosin and reflecting the formation or dissolution of apical actomyosin foci. Previous measurements  have shown that myosin and actin concentrations are proportional throughout the AS oscillation cycle \cite{Blanchard2009}. We thus assume a first-order binding kinetics of myosin to the F-actin cortex,
\be\label{eq:myo_con}
\frac{dM_k}{dt} = k_{on} A_k - k_{off} M_k\,,
\ee
where  $A_k$ and $M_k$ are the F-actin and myosin concentrations of cell $k$ respectively, and $k_{on}$ and $k_{off}$ are the myosin binding and unbinding rates. Live-imaging of actomyosin and membrane dynamics of AS cells has also revealed that actomyosin peaks precede the troughs in cell area in AS oscillation cycles. In an analogous formulation to the single cell cortical dynamics of  \cite{Sedzinski2013}, we then take actin behaviour to be given by
\be\label{eq:actin_con}
\frac{dA_k}{dt} = \frac{1}{\tau}\left(S_k \,a_0 - A_k\right)\,,
\ee
where $\tau$ is the medial F-actin turnover time. Crucially, the regulation of actin depends on a reference actin density $a_0$ and the apical surface area $S_k$. 
Assuming the on-off myosin kinetics is much faster than the typical AS oscillation timescale \cite{Kovacs2007,Harris2013}, we can approximate the myosin equation by its steady state solution $M_k \simeq \left(k_{on}/k_{off}\right) A_k$. Substituting this into \eqref{eq:motion} and rewriting our equations in terms of the actin density $a_k \equiv A_k/S_k$ yield
\be\label{eq:motion_act}
\eta \frac{d\vec{x}_i}{dt} = \sum_{\rm k} \vec{u}_{ik}\left[ \kappa \left(l_{ik}-l_{ik}^{0}\right)+\alpha_m\,f_m\left(\tfrac{a_k}{a_0}\right)\right] \,,
\ee
\be\label{eq:actin}
\frac{da_k}{dt} =\frac{1}{\tau}a_0 - \left( \tfrac{\dot{S}_k}{S_k} + \frac{1}{\tau} \right) a_k\,,
\ee
which, together with \eqref{eq:ratchet}, are the equations implemented in our simulations.

Laser ablation experiments \cite{Jayasinghe2013} and computer simulations \cite{Wang2012} suggest AS cells are under low elastic strain, and we therefore set the starting spoke rest lengths to their initial lengths in the segmentation. The initial actomyosin concentration of individual cells is taken from a uniform distribution within 10\% of the steady state concentration. In order to simulate tissue dynamics, we have adopted two different tissue configurations. For an initial exploration of the parametric dependence of our model, we have first used a hexagonal lattice of $90$ with identical spoke rest lengths $l_0$. For subsequent simulations, we have used a snapshot of an AS segmentation during early DC as our initial tissue configuration (Fig.~\ref{fig:forces}), where we connect the anteriormost and posteriormost AS cell vertices to external fixed vertices representing the AS canthi. We restrict the vertices-canthi forces to a purely passive force, with rest length set to the initial vertex-canthus distance and remaining constant throughout DC. Apart from the external canthi, all vertices move according to \eqref{eq:motion_act}. Lastly, as the final stages of DC involve contributions from zippering and the actin cable \cite{Kiehart2000,Gorfinkiel2011} which are not present in our model, we restrict our tissue simulations to the early and slow DC phases only. Simulations were implemented via a custom java-based ODE solver with the fourth-order Runge-Kutta method, and subsequent analysis and fit of model parameters was done using Mathematica 9.0 (Wolfram Research, Champaign, IL) and R \cite{R:main}.

\section{Results}
\label{results}

\begin{figure}
\begin{subfigure}{0.48\textwidth}
  \includegraphics[width=\textwidth]{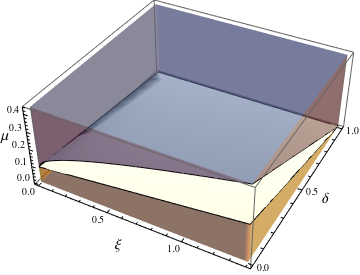}
  \caption{} \label{fig:1a}
\end{subfigure}
\begin{subfigure}{0.48\textwidth}
  \par\medskip 
  \includegraphics[width=\textwidth]{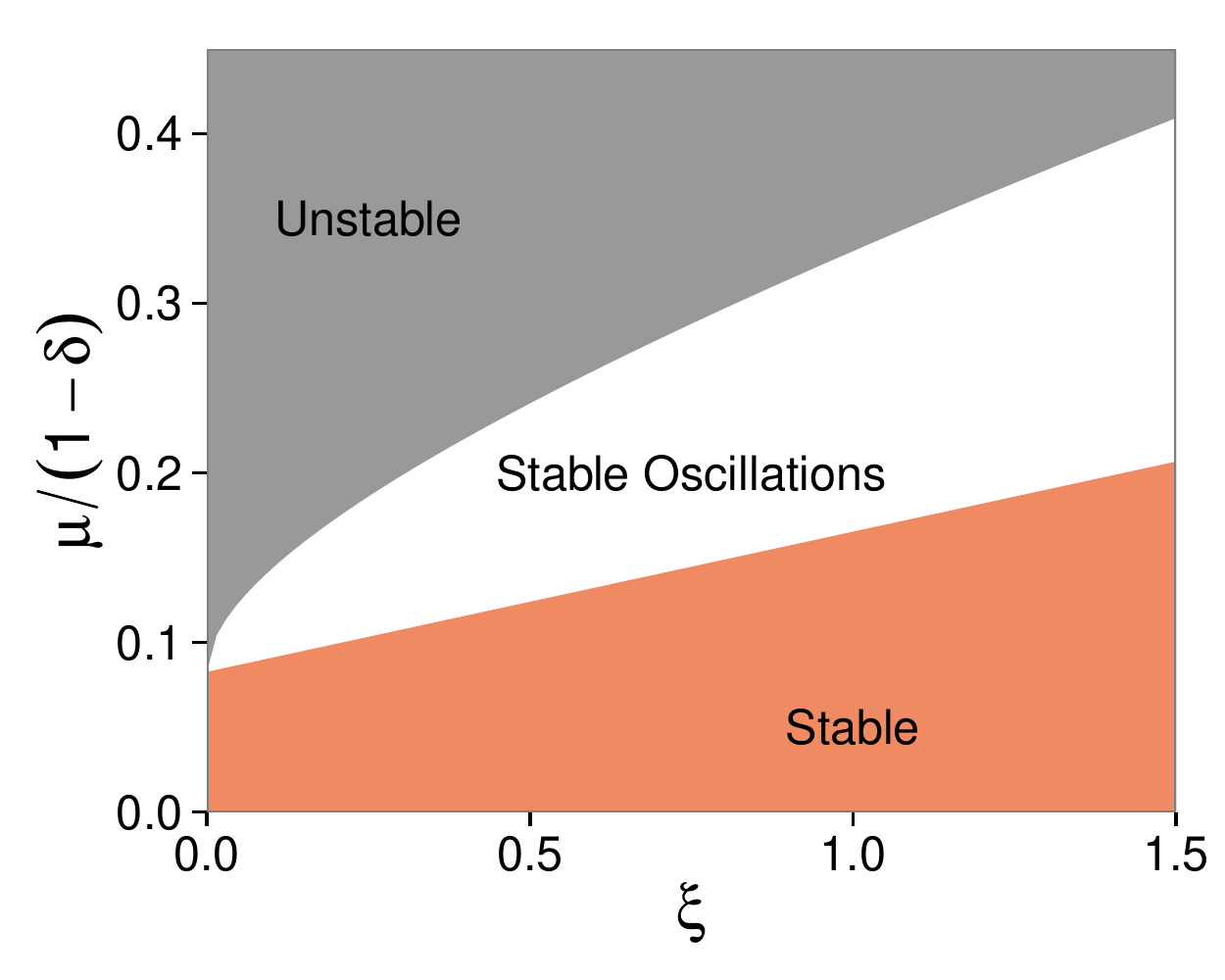}
  \caption{} \label{fig:1b}
\end{subfigure}
\caption{\label{fig:phase_sc} {\footnotesize Phase diagram of a linear stability analysis of a single-cell implementation of our model. We observe a regime of stable oscillations for small values of the dimensionless active force coefficient $\mu$ and for most values of the dimensionless actomyosin turnover rate $\xi$. The effect of the net contraction of the spoke rest length $\left(1 - \delta) l_0(0\right)$ is to gradually reduce the space of stable oscillations.}}
\end{figure}

We have investigated the oscillatory dynamics of our model by first considering a single-cell implementation of our equations. The results of a linear stability analysis are illustrated in Fig.~\ref{fig:phase_sc}, where we observe a regime of stable oscillations of cell area and actomyosin density. The behaviour of our single-cell model depends on three quantities, the ratio of the viscoelastic relaxation time to the actin turnover time $\xi \equiv \eta/\kappa\tau$, which defines a dimensionless actin turnover rate, the ratio of the active actomyosin force to the passive viscoelastic resistance $\mu \equiv \alpha_m/\kappa l_0$, which defines a dimensionless active force coefficient, and the spoke rest length reduction relative to the original rest length $\delta \equiv t\,\nu \eta/\kappa l_0$, which defines a dimensionless ratchet coefficient\footnote{Strictly speaking, the time-dependence of $\delta$ implies steady state solutions for this system only exist for $\nu = 0$. In order to perform the stability analysis, we thus pick fixed values of $\delta$, ignoring its time-dependence, and consider the behaviour of the system around $\hat{l}_0 = (1 -\delta) l_0(0)$. }. We note that the effect of the net area reduction is to gradually reduce the space of sustained oscillations (Fig.\ref{fig:phase_sc}a).

\begin{figure}
\begin{subfigure}{0.48\textwidth}
  \includegraphics[width=\textwidth]{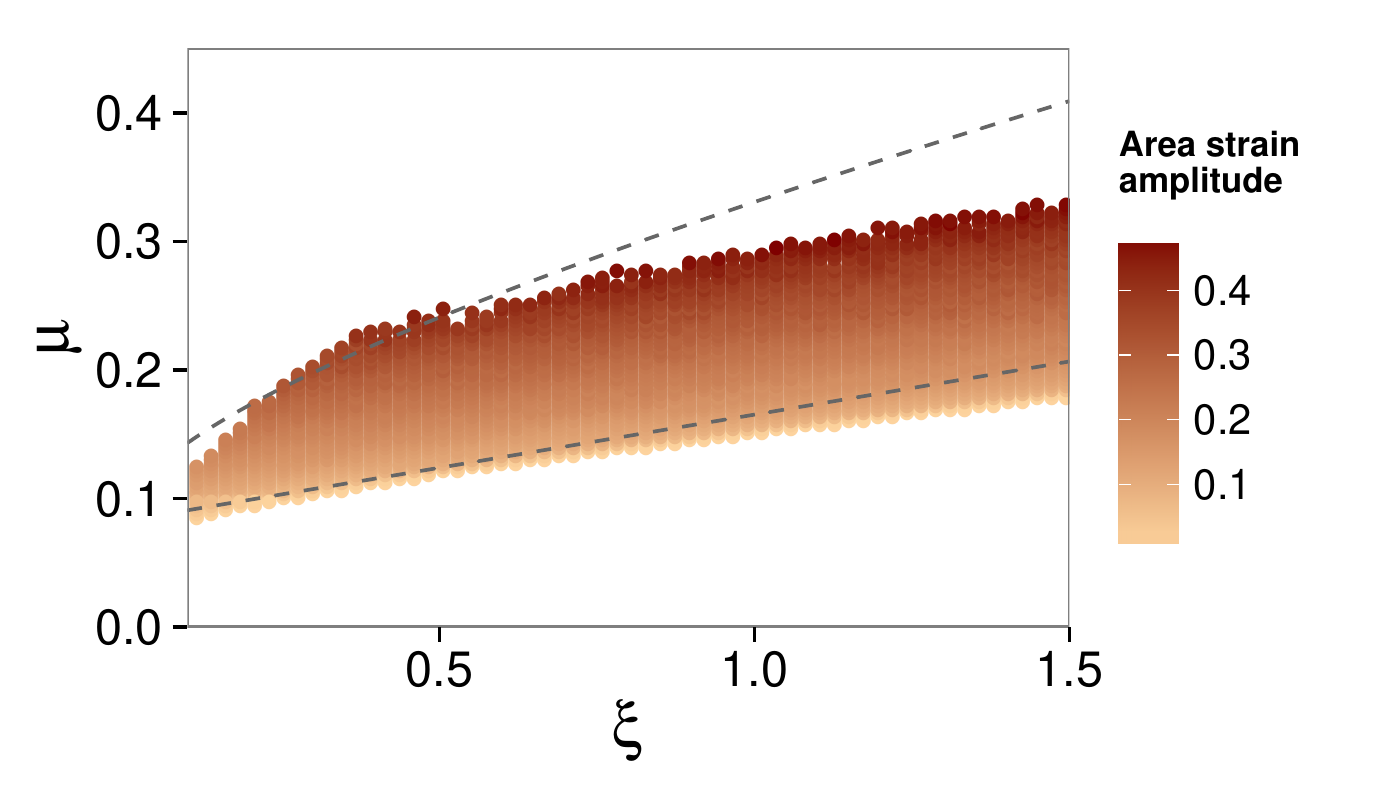}
  \caption{} \label{fig:1a}
\end{subfigure}
\begin{subfigure}{0.48\textwidth}
  \par\medskip 
  \includegraphics[width=\textwidth]{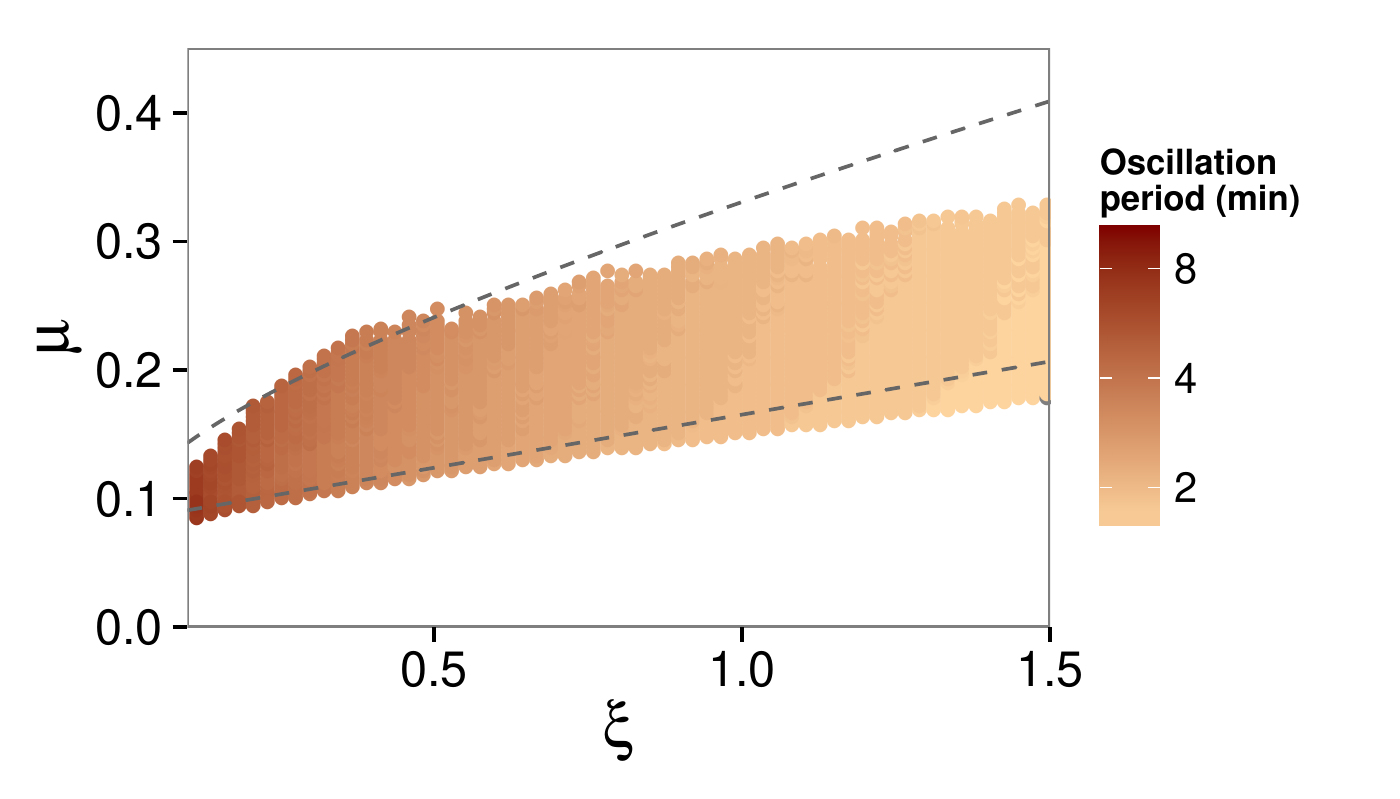}
  \caption{} \label{fig:1b}
\end{subfigure}
\begin{subfigure}{0.48\textwidth}
  \includegraphics[width=\textwidth]{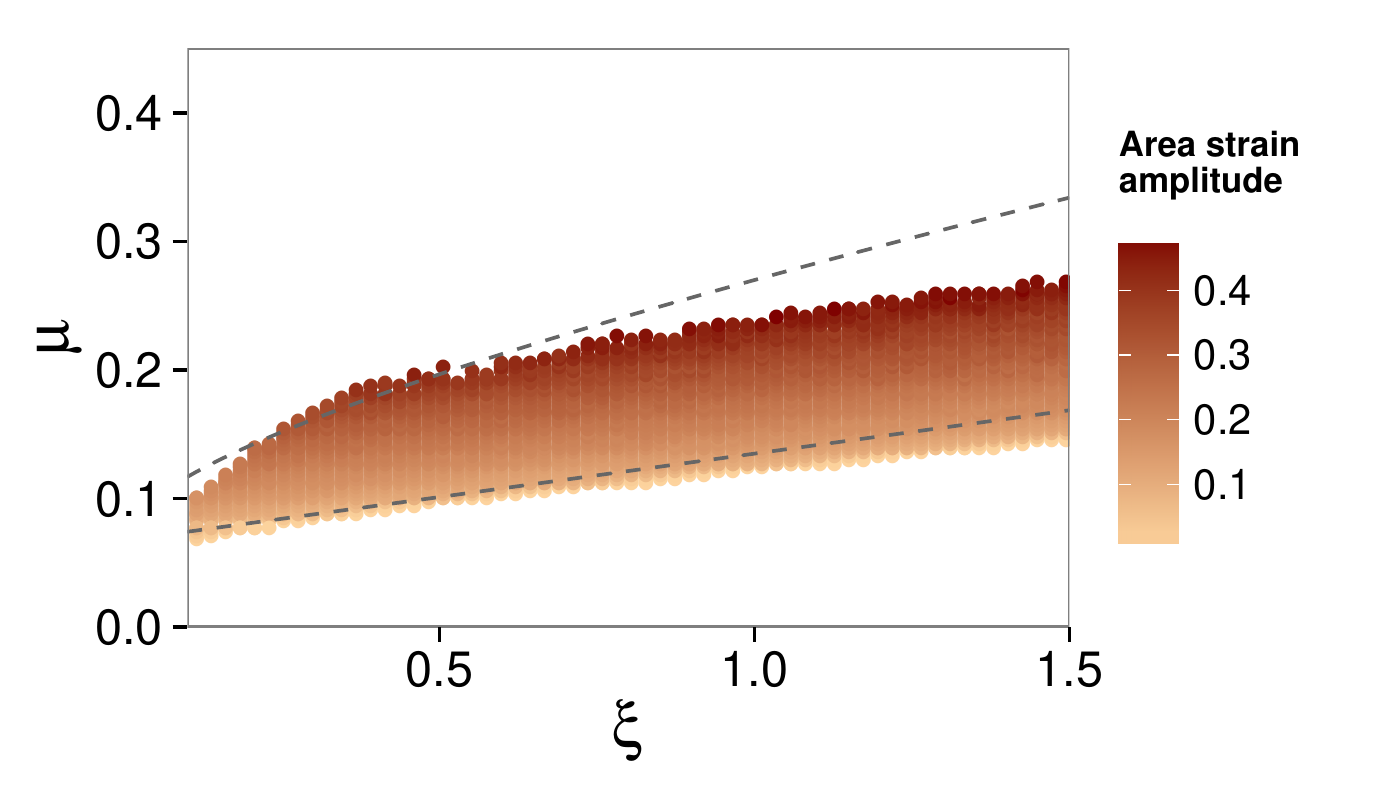}
  \caption{} \label{fig:1c}
\end{subfigure}
\begin{subfigure}{0.48\textwidth}
  \par\medskip 
  \includegraphics[width=\textwidth]{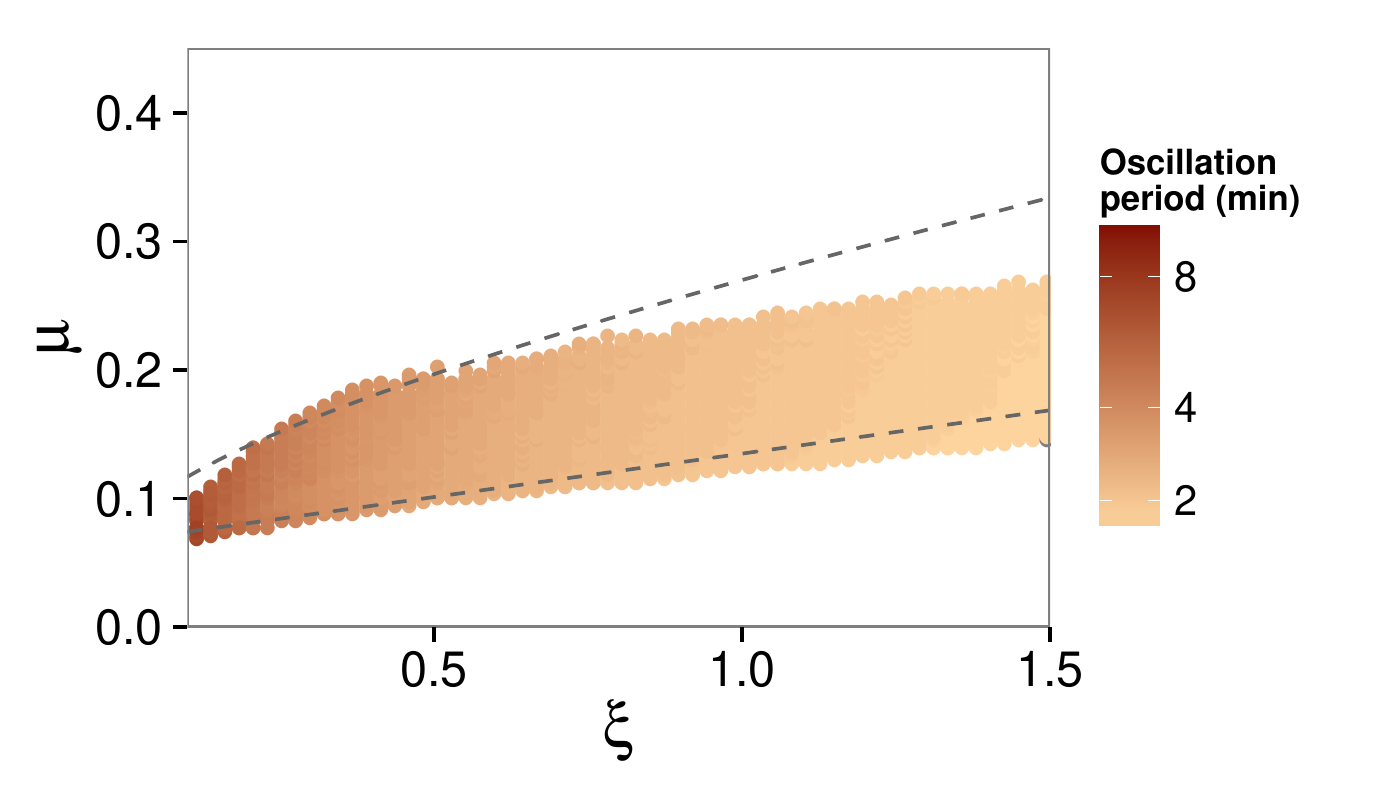}
  \caption{} \label{fig:1d}
\end{subfigure}
\begin{subfigure}{0.48\textwidth}
  \includegraphics[width=\textwidth]{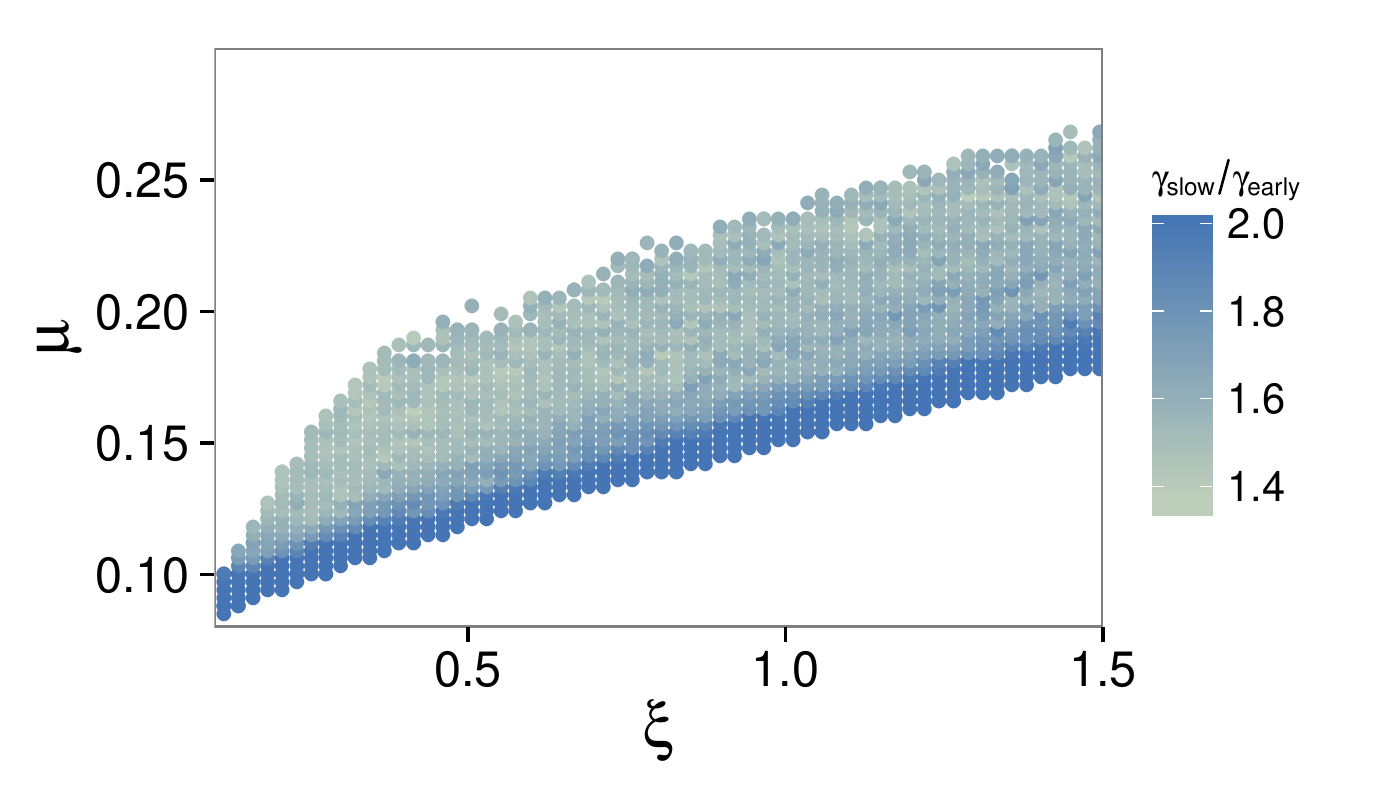}
  \caption{} \label{fig:1e}
\end{subfigure}
\begin{subfigure}{0.48\textwidth}
  \includegraphics[width=\textwidth]{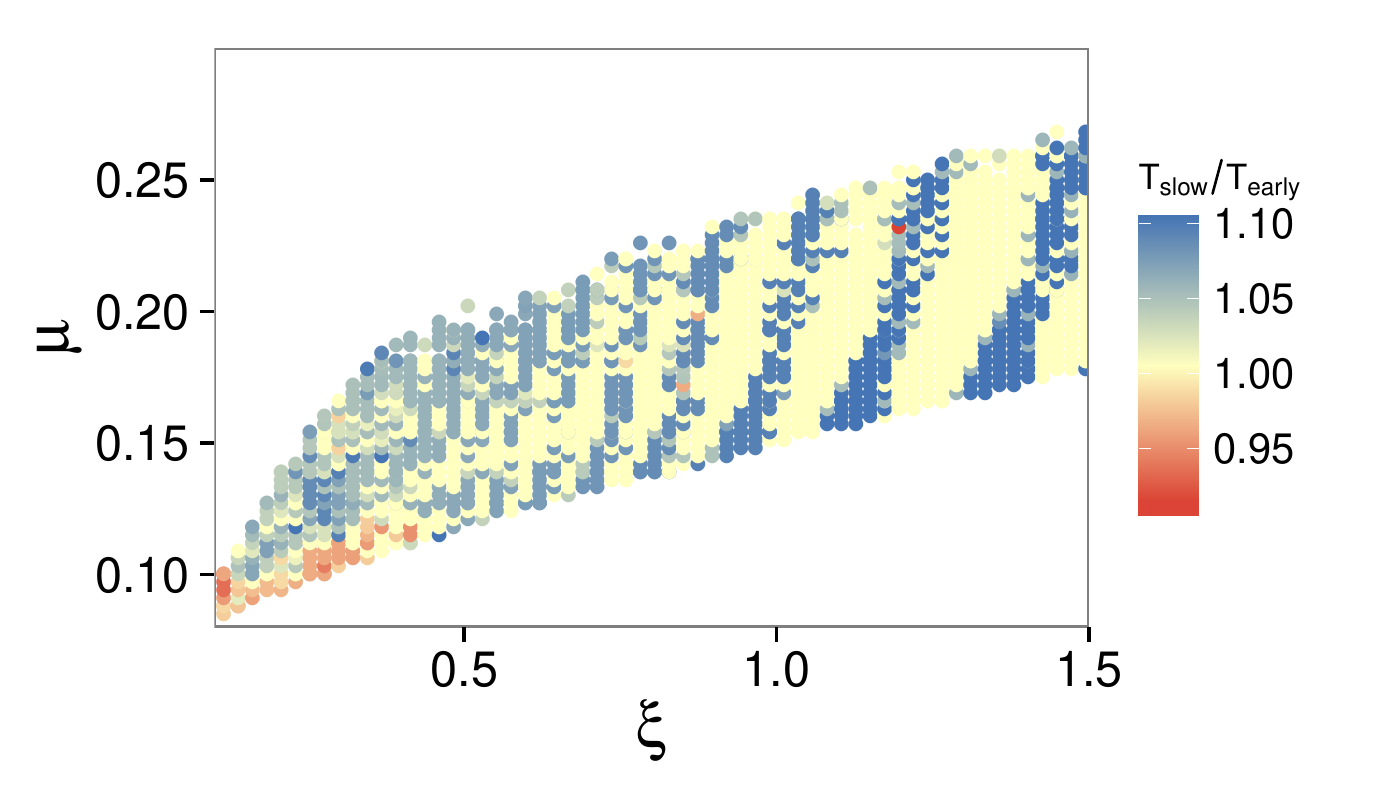}
  \caption{} \label{fig:1f}
\end{subfigure}
\caption{\label{fig:hex_scan} {\footnotesize Parameter dependence of AS oscillations. Area strain amplitudes and oscillation periods of early (a, b) and late slow (c,d) DC phases obtained via parameter scans; regions bounded by the dashed lines correspond to the regime of stable oscillations for the single cell implementation of the model. Ratios of strain amplitude (e) and oscillation period (f) between slow and early simulations run under the same parameter values.}}
\end{figure}

Motivated by these findings, we numerically investigated the existence and resulting dynamics of the oscillatory regime at the tissue level via a parameter scan. In order to reduce cell-to-cell variability and increase the stability and accuracy of our parameter scan, we performed these simulations on a hexagonal lattice of 90 cells with identical spoke rest lengths $l_0$ and with initial actomyosin density uniformly distributed around 10\% of the equilibrium density $a_0$. Setting the net area reduction rate to $\nu\,(t \geq 0) = 3.8 \times 10^{-5} \mu m \,s^{-1}$ so that cells reach a rest area of $2/3$ of their original rest area by the end of slow DC and letting the other parameters vary, we have determined the region of stable oscillatory behaviour. Within this region, we have then investigated the parametric dependence of the period and amplitude of area oscillations. The results are shown in Fig.~\ref{fig:hex_scan} for early DC ($t < 0\,s$) and for the final portion of slow DC ($t > 0$ and $S_{\rm rest}(t) \approx \frac{2}{3}\,S_{\rm rest}(0)$).

We observe that, although the phase region of stable oscillations for the single-cell and multicellular implementations of the model are not identical, there is a notable overlap between the two (Figs.~\ref{fig:hex_scan}a,c), particularly for low values of $\xi$. This suggests that the effect of the mechanical coupling among neighbouring cells is to modulate the dynamics of oscillations which may already be present at the single-cell level. Comparing the size of the oscillatory regions between slow and early DC in our parameter range, we can see that, in a similar manner to the single-cell case, one effect of the area ratchet is to restrict the space of stable oscillations. Focusing on the two main quantitative features of AS oscillations, we observe a strong dependence of area strain amplitude on the dimensionless active force coefficient, with increasing amplitude for increasing values of $\mu$ (Figs.~\ref{fig:hex_scan}a,c), and a strong dependence of oscillation period on the dimensionless actomyosin turnover rate, with increasing period for decreasing $\xi$ (Figs.~\ref{fig:hex_scan}b,d). These parametric dependences are not unexpected, as higher values of $\mu$ correspond to an increase in the relative active force magnitudes and hence an increase in cell deformation, while higher values of $\xi$ imply both a faster turnover of actomyosin around the reference density $a_0$, which also represents the threshold for active force activation, and hence faster expansion and contraction half-cycles. 

Published experimental data has established that, as DC progresses, AS cell oscillations decrease in both amplitude and period \cite{Blanchard2009,David2010,Sokolow2013}, with average strains going from ca. 9\% to ca. 5\% between the early and fast phases, and average oscillation cycles going from ca. 230s to ca. 150s. The time evolution of AS oscillation properties is a central biological observation that is unaccounted for in current models of DC. While the internal ratchet implementation in the model proposed by Wang et al \cite{Wang2012} leads to a reduction of the absolute amplitude of oscillations, it is not clear whether this mechanism also leads to an attenuation of strain amplitude, as the ratchet will also reduce the cell rest area. We thus investigated how the evolution of AS oscillations could arise in our model. From our parameter scan we observe that, for fixed values of $\mu$ and $\xi$, the area ratchet effects a relative change in AS strain amplitude and period between early and slow DC (Figs.~\ref{fig:hex_scan}e,f), and we then asked whether the ratchet could account for the observed oscillation evolution. The biological values of strain amplitude and period for early DC reported in the literature coupled with the parameter dependence of our model indicate that the range of physiological values for our parameters should be restricted to the lower left-hand region of our parameter space ($\mu \lesssim 0.2,\, \xi \lesssim 0.5$). Focusing on this region, however, we note that the ratchet causes an increase in both amplitude and period of oscillations for most fixed values of our parameters, indicating that our ratchet cannot account for AS oscillation evolution. In order for our model to reproduce the experimental data and given the observed parametric dependence of our oscillations, we thus require that the parameters $\mu$ and $\xi$ evolve as DC progresses. Choosing initial values for our parameters $\left\{\mu,\xi\right\}_{\rm early}$ compatible with early DC oscillations, we then performed a sweep for parameter values $\left\{\mu,\xi\right\}_{\rm slow}$ which recapitulated slow DC oscillations. Comparing the parameter values compatible with early and slow DC and computing their relative differences, we find that, in order to capture DC evolution dynamics, our model requires both $\mu$ and $\xi$ to increase with time.

\begin{figure}
\includegraphics[width=\textwidth]{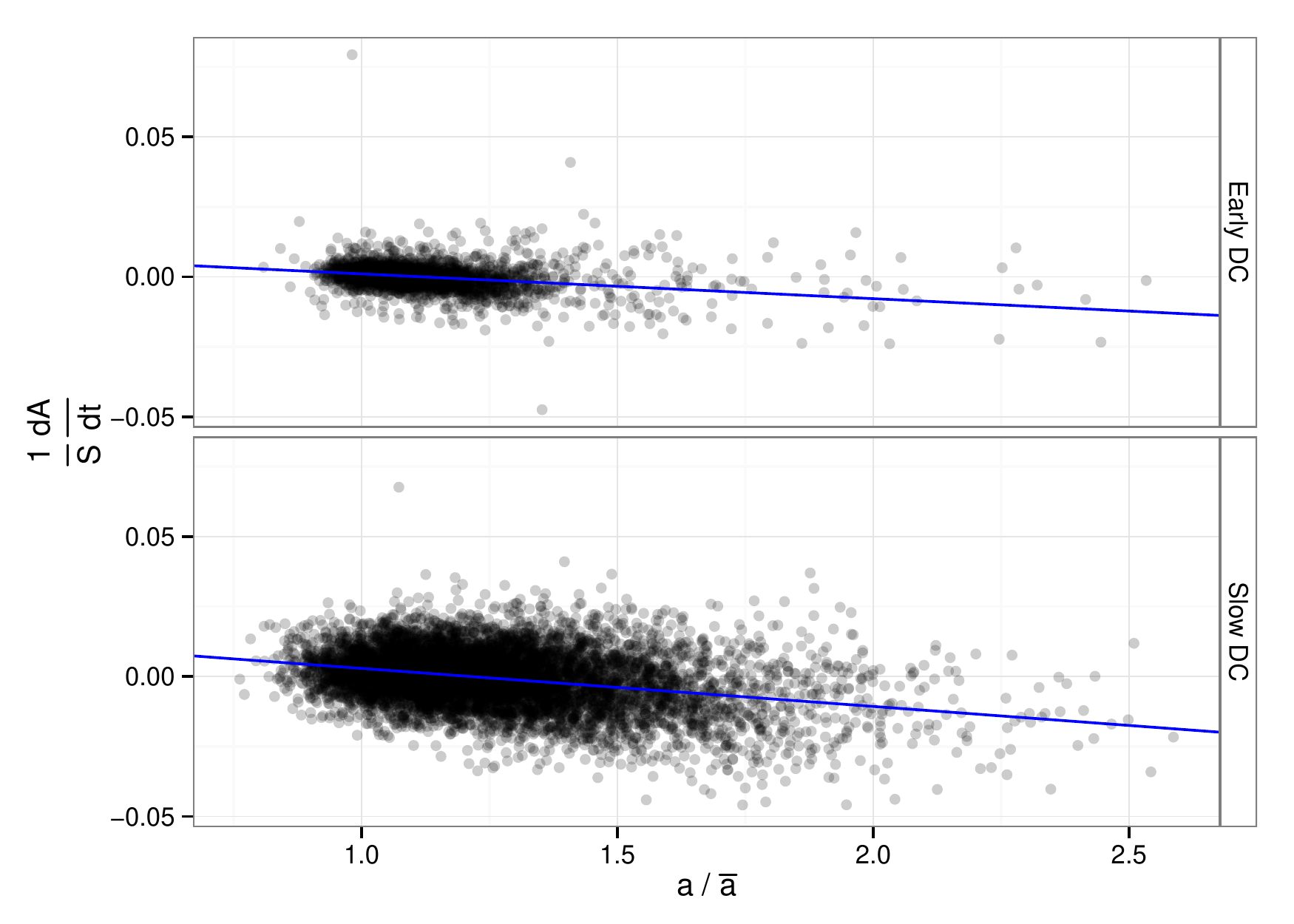}
\caption{\label{fig:fit} {\footnotesize Fit of AS cell area and actin fluorescence intensity (sGMCA) measurements (black points) to the actomyosin equation in the model, yielding the actomyosin turnover rate (blue line); actin fluorescence density $a$ has been rescaled by a long-timescale trend in fluorescence density $\bar{a}$ in order to account for background fluorescence and photobleaching effects. We observe an increase in the turnover rate between early and slow DC.}}
\end{figure}

Under the assumption that the viscoelastic relaxation timescale of AS cells does not change over time, an increase in $\xi$ is equivalent to an increase in the cortical actomyosin turnover rate of AS cells. In order to determine whether this increase occurs in vivo, and thus whether the requirement of our model is biologically sound, we performed live imaging of early and slow wild-type embryos carrying a membrane marker (ArmadilloYFP) and an actin reporter (sGMCA \cite{Kiehart2000}) and extracted measurements of AS cell area and actin fluorescence intensity (Movie S1). Using a generalized linear mixed model \cite{Pinheiro:2009}, we then fitted this data to our equation for actomyosin dynamics, Eq.~\eqref{eq:actin}, yielding estimates of the bulk apical actin turnover rate of AS cells in early and slow DC. A parametric bootstrap test \cite{pbkrtest} shows that the resulting turnover rates are dependent on developmental phase ($p  = 0.018$), and the values of the parameter fit for early and slow DC are shown in Table \ref{t.1} and illustrated in Fig.~\ref{fig:fit}. In agreement with the requirement of our model, we find that actin turnover rate increases as DC progresses, with turnover half-lives going from $\tau_{\rm early} = 112.9 \pm 7.1s$ to $\tau_{\rm slow} = 73.6 \pm 2.5s$. We note that these rates correspond to the bulk turnover of the entire apical actin network of individual cells.

For fixed passive viscoelastic coefficients, an increase in $\mu$ implies an increase in the active contractile force per relative unit myosin. Although we have not verified if this increase obtains form biological DC measurements, such an increase is biologically plausible, and we speculate that it might be caused by changes in the actin network architecture during DC development (see Discussion).

\begin{table}
\begin{center}
\begin{tabular}{lllllll}
DC Phase &  1/$\tau\,(s^{-1})$ \qquad& $N_{\rm cells}$ \qquad & $N_{\rm embryos}$ \\ \hline
Early & $8.85 \pm 0.59 \times 10^{-3} $ & 34 & 4  \\
Slow & $13.58 \pm 0.47 \times 10^{-3} $ & 39 & 8  \\
\end{tabular}
\end{center}
\parbox[c]{\textwidth}{\caption{\label{t.1}{GLMM fit of actomyosin turnover rate for DC phases. }}}
\end{table}

\begin{figure}
\begin{subfigure}{0.48\textwidth}
  \includegraphics[width=\textwidth]{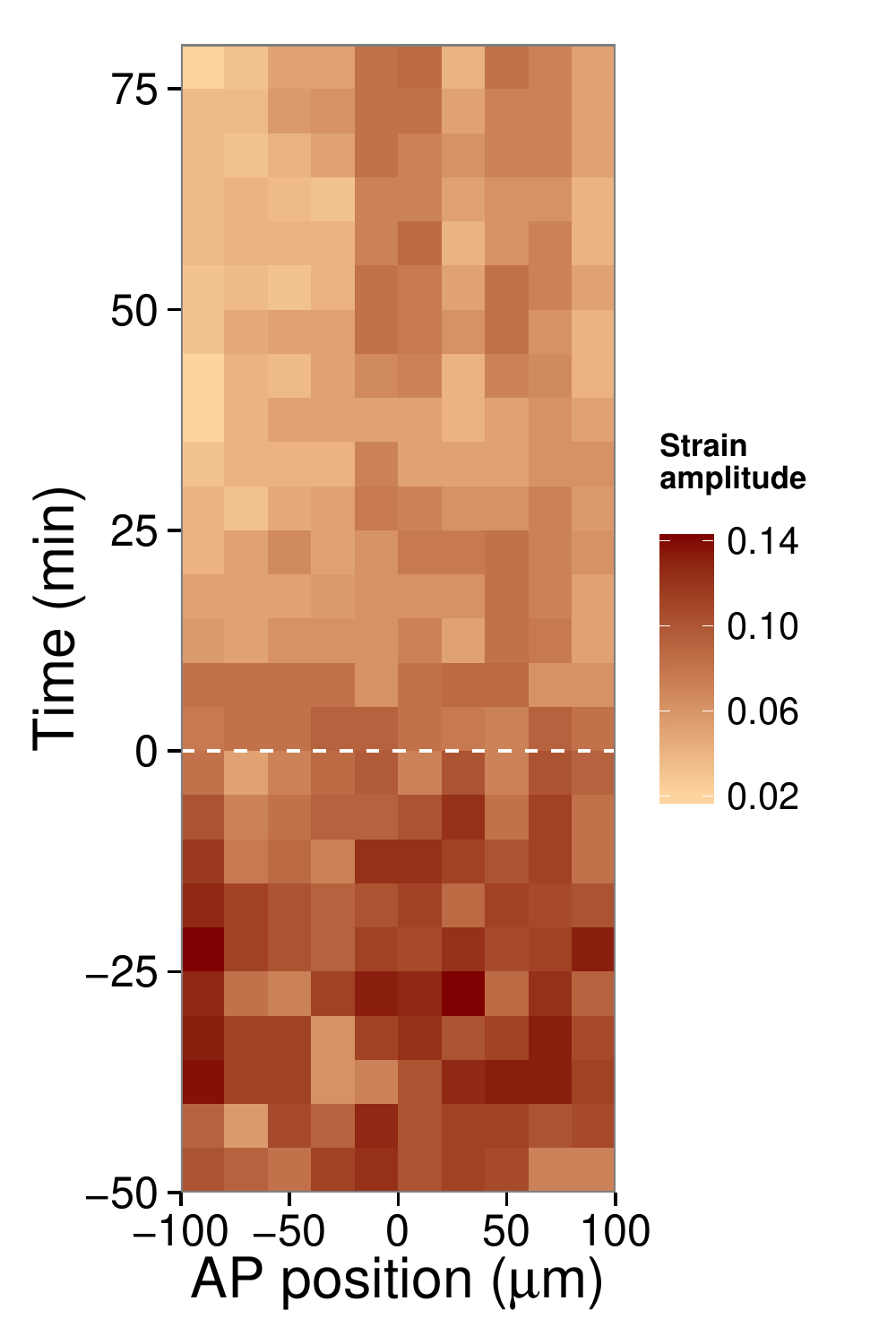}
  \caption{} \label{fig:1a}
\end{subfigure}
\begin{subfigure}{0.48\textwidth}
  \includegraphics[width=\textwidth]{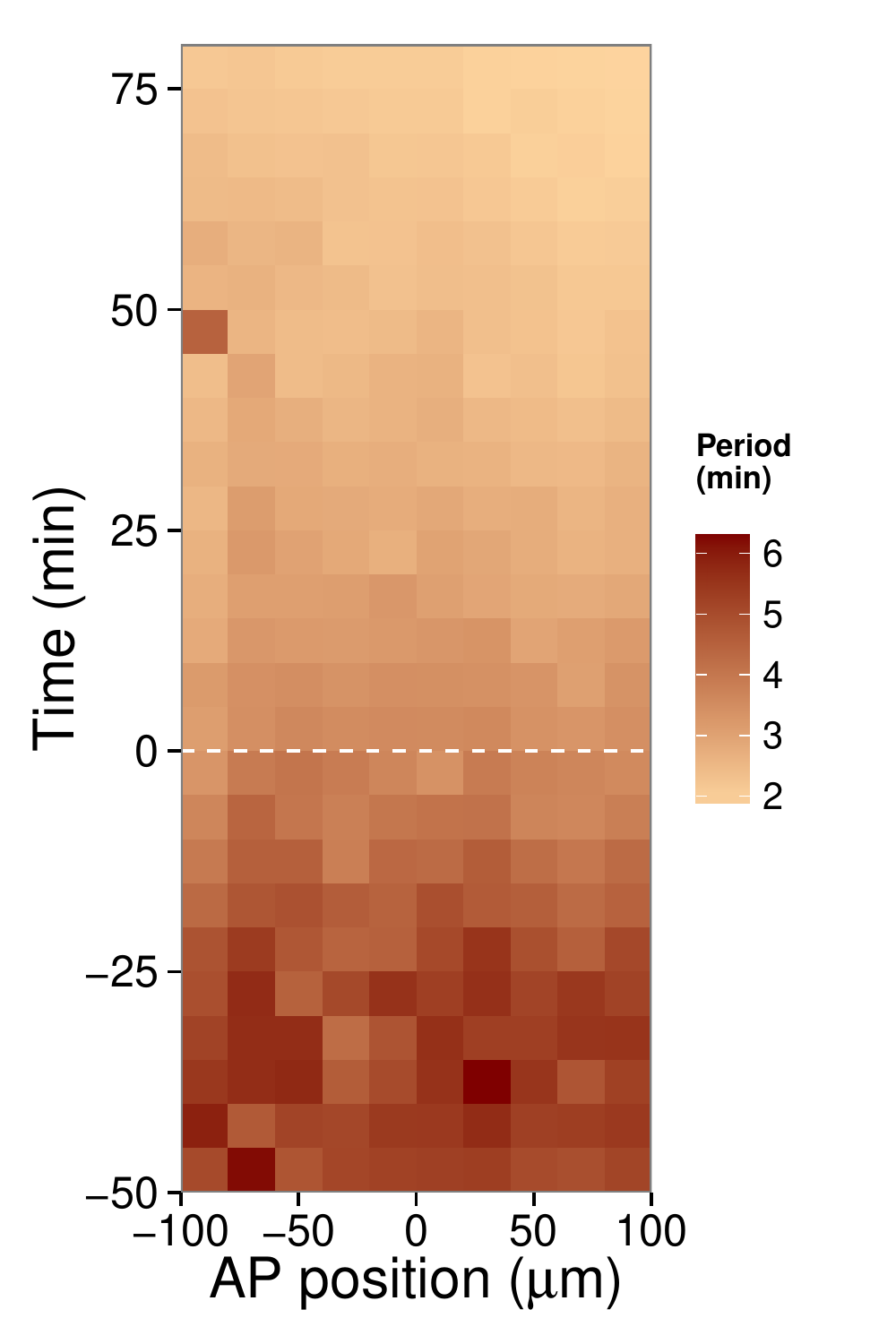}
  \caption{} \label{fig:1b}
\end{subfigure}
\caption{\label{fig:seg_evo} {\footnotesize Temporal evolution of the (a) strain amplitude and (b) period of AS oscillations in a tissue simulation. Time $t = 0$ min corresponds to the onset of slow DC and anteriormost cells correspond to negative values of AP position.}}
\end{figure}

\begin{figure}
\begin{subfigure}{0.64\textwidth}
  \includegraphics[width=\textwidth]{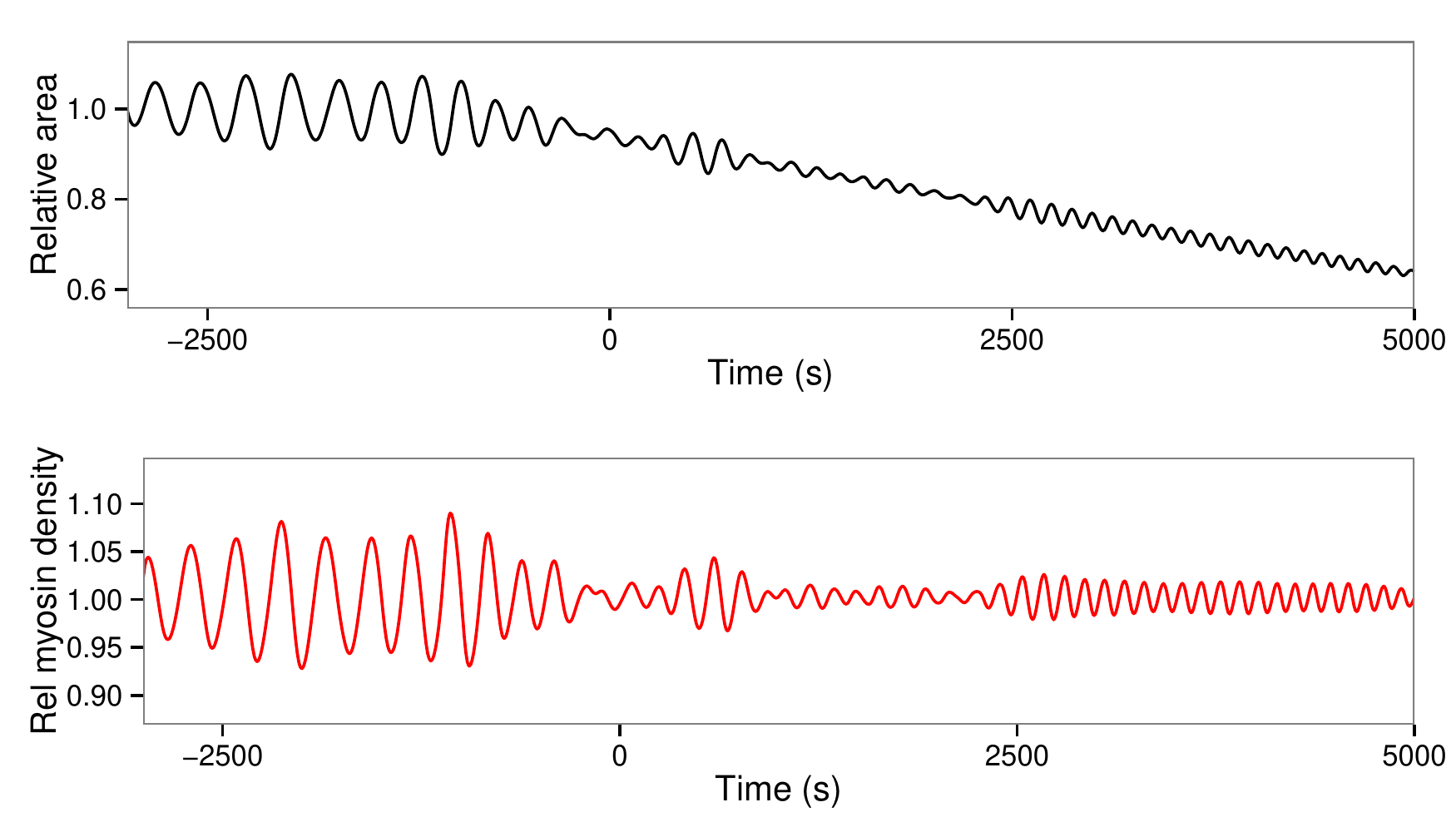}
  \caption{} \label{fig:1a}
\end{subfigure}
\begin{subfigure}{0.32\textwidth}
  \includegraphics[width=\textwidth]{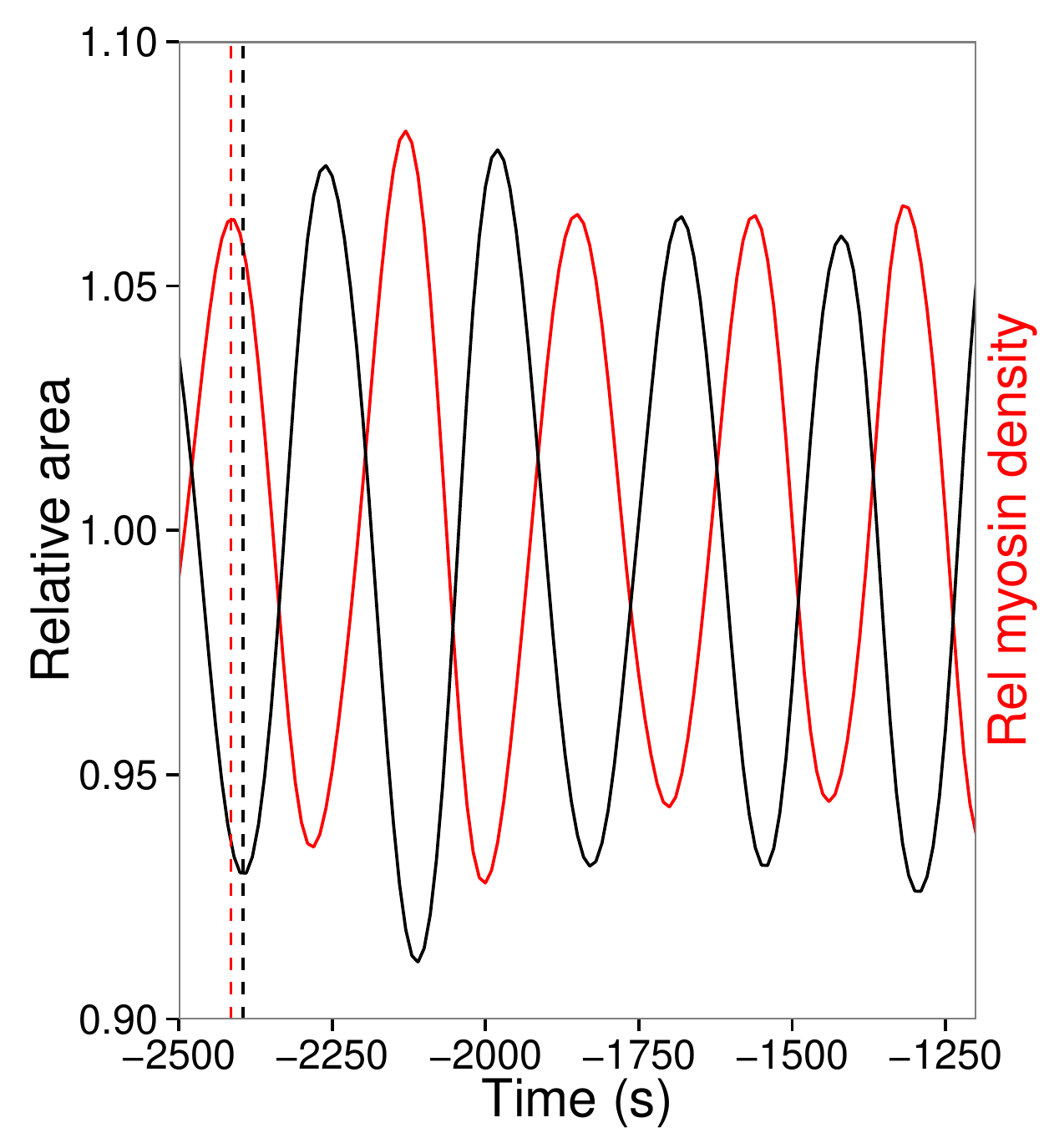}
  \caption{} \label{fig:1b}
\end{subfigure}
\caption{\label{fig:seg_eg} {\footnotesize Behaviour of a sample cell in AS tissue simulation. (a) Normalized area and normalized actomyosin density; cells oscillate around fixed rest area until the beginning of the slow phase ($t = 0\,s$) when an internal ratchet is activated via the linear reduction of spoke rest lengths. (b) Cell area (black) and actomyosin density (red) oscillate in anti-phase, with peaks in actomyosin density (red dashed line) preceding troughs in area (black dashed line).}}
\end{figure}

Guided by these considerations, we have thus conducted simulations of DC tissue dynamics in which both $\xi$ and $\mu$ increase with time. We have used the AS segmentation in Fig.~\ref{fig:forces} as our starting tissue configuration and selected initial and terminal values of $\xi$ and $\mu$ from our parameter scan so as to reproduce the evolution of AS amplitude and period observed in vivo. These parameters are kept fixed to their initial values $\left\{\mu,\xi\right\}_{\rm early}$ throughout the early phase ($-3000s \leq t < 0s$) and start linearly reducing at the onset of the slow phase ($t = 0s$), until reaching their terminal values $\left\{\mu,\xi\right\}_{\rm slow}$ at the end of the slow phase ($t = 4800s$). We implement the net area ratchet by letting $\nu\,(t<0) = 0$ for the slow phase and then setting $\nu\,(t \geq 0) = 3.8 \times 10^{-5} \mu m \,s^{-1}$ so that cells have on average reduced their rest area by a third of their original area by the end of the slow phase. Initial conditions were set by first defining the equilibrium configuration as the one corresponding to the segmented image with equal values of actomyosin density $a_0$ for all cells, and setting the starting values of actomyosin density for individual cells to be uniformly distributed around 10\% of the equilibrium configuration. The results of our simulation (Movie S2) are shown in Fig.~\ref{fig:seg_evo} for the whole tissue and in Fig.~\ref{fig:seg_eg} for an example cell. We note that cell area and actomyosin density oscillate in anti-phase, with peaks in actomyosin preceding the troughs in cell area (Fig.~\ref{fig:seg_eg}b), in agreement with experimental observations \cite{Blanchard2009}. As a consequence of the parameter evolution, the period and amplitude of oscillations reduce over time in accordance with the biological data (Fig.~\ref{fig:seg_evo}). Taken together, these results show our model recapitulates DC dynamics at the tissue and cell level and suggests an alternative, cell-autonomous mechanism for the emergence of oscillations.

\section{Discussion}
\label{discussion}

We have presented here a cell-level model of DC that reproduces AS cell oscillatory and contractile behavior during all but the fast phase of dorsal closure. This model represents an alternative framework to the one presented by Wang and colleagues \cite{Wang2012}, in which cell shape oscillations emerge from cell-autonomous feedback loops between a putative signal that activates Myosin and is degraded by Myosin activity and mechanical coupling between cells. Our model, inspired by a previous model of single cell oscillations during cytokinesis \cite{Sedzinski2013}, shows that sustained cell oscillations can also emerge in sillico through the coupling of myosin-driven active forces, actomyosin turnover and cell deformation. In our model, the parameter space in which both single cell oscillations and multicellular oscillations arise shows significant overlap, and this is particularly true for the parameter region which best reproduces experimental data. Our model thus suggests that cell oscillations are a cell-autonomous phenomenon, in agreement with previous experimental results \cite{Jayasinghe2013} and in contrast to early theoretical models of AS cell behaviour \cite{Solon2009}.

Using our model, we have investigated the temporal evolution of the amplitude and period of cell area oscillations. A key prediction of our model is that, in order for amplitude and period to reproduce the experimental observations, the ratio of the viscoelastic relaxation time to actomyosin turnover time as well as the ratio of contractile force per unit Myosin to passive viscoelastic resistance have to increase over time. The predicted increase in the former ratio suggests that the actin cortex turnover is a key parameter controlling cell oscillations. This suggestion is further strengthened by a fit of the actomyosin equation in our model to live-imaging data, which shows that the bulk turnover time of the actin cortex decreases during the transition from the early phase to the slow phase of DC. It has been proposed that turnover dynamics of the actomyosin cortex could strongly affect cell shape changes and cell mechanical properties \cite{Salbreux2012} The turnover of the actin cortex can be regulated by actin-binding proteins, including Myosin and other actin cross-linkers. Interestingly, Myosin activity has been shown to promote actin network disassembly \cite{Wilson2010} and accelerate actin turnover \cite{Guha2005} in other contexts, while cross-linkers such as alpha-actinin have the opposite effect, slowing down actin turnover \cite{Mukhina2007}. Thus, subtle changes in the activity of these actin-binding proteins could modulate oscillation dynamics. Our preliminary results of FRAP experiments measuring the turnover of Myosin in AS cells also suggest that turnover rates of the actomyosin cortex accelerate during DC (unpublished observations). 

While we have not verified whether an increase in the ratio of contractile force per unit Myosin to passive viscoelastic resistance obtains in vivo, we speculate this might be caused by an increase in the Myosin contractile force. Myosin contraction force has been suggested to depend on the architecture of the actin network \cite{Reymann2012,Alvarado2013}. In particular,  in vitro experiments on reconstituted networks have shown that Myosin force is modulated by the ratio of branched to bundled actin structures, with an increase of force for increasing bundling \cite{Reymann2012}. In addition, experiments on cross-linked actin networks suggest that, in order for Myosin motors to coordinate contractions at long lengthscales and prevent local rupturing of the network, the ratio between motor and cross-linker concentrations should be balanced \cite{Alvarado2013}. Thus, an interesting hypothesis that arises from our model is that the increase in Myosin force could be due to a change in the proportion of branched to bundled actin networks, a change in the cross-linker concentration, or both. Interestingly, in the context of DC, we have observed  that promoting the overbranching of actin networks in the AS through the ectopic expression of a constitutively active form of Wasp prevents AS cell contractions (unpublished). 

In the model of Wang et al \cite{Wang2012}, the period of oscillations is determined by the association rate of myosin in response to the putative activating signal, with an increase in the association rate leading to shorter periods. While Wang et al keep the oscillation period fixed throughout DC in their simulations, allowing this period to evolve to match the experimental observations would thus presumably imply an increase in the myosin association rate. It is worth noting that such an increase might be connected to both the increase in actomyosin turnover time and Myosin contractile force predicted in our model. Recent measurements of AS cell behavior \cite{David2013} have shown that interactions between signalling molecules of the Par complex, which are regulators of AS actomyosin activity \cite{David2010}, increase as DC progresses, suggesting another molecular underpinning to the phenomenological evolution of AS actomyosin dynamics predicted by our model.

Although our model is able to quantitatively reproduce AS cell behaviour during DC, it crucially relies on the assumption of the existence of a reference actin density that cells tend to spontaneously maintain. It is the interplay between fluctuations around this reference density driven by cortex turnover and cell shape deformations that drives sustained oscillations. While it has been proposed in the context of cleavage furrow constriction during cytokinesis that the number of actomyosin contractile units scale with the size of the contractile ring \cite{Carvalho2009}, which would suggest an equilibrium actomyosin density, to our knowledge there is no other experimental evidence for such an equilibrium state of the actin cortex nor how it could be maintained. However, the cytoskeleton is the primary mechanosensing system of the cell and it is thus possible that such equilibrium states exist. 

Another feature of the model is that actomyosin and cell area oscillations are tightly coupled and the observed decrease in amplitude of cell area oscillations goes hand in hand with a decrease in amplitude of actomyosin oscillations. Our preliminary results on the evolution of the amplitude of Myosin oscillations (unpublished observations) suggest that this might not be the case. This decoupling between actomyosin activity and cell area deformation has also been observed in apically contracting cells during gastrulation both in \textit{C. elegans} and \textit{Drosophila} and has led to the suggestion that apical cell contraction arises from linking cell contacts to a preexisting contractile actomyosin network \cite{RohJonson2012}. Altogether, these results suggest to us that actomyosin oscillations may emerge from an autonomous chemical feedback mechanism at the level of single cells. The mechanical coupling between cells would then modulate the oscillatory behaviour and would contribute to the coordination of the pulsatile contractile behaviour across the whole tissue.

\section{Acknowledgements}
We thank the following funding bodies for their support: Herchel Smith Fund (PFM), Biotechnology and Biological Sciences Research Council (GBB), Spanish Ministry of Science (NG and JD, grant BFU2011-25828), ÒRam\'{o}n y CajalÓ fellowship award (NG), Marie Curie Career Integration Grant (NG, PCIG09-GA-2011-293479) and the European Science Foundation (NG, QuanTissue grant 4882 ). We thank Sabine Fischer for the segmented AS tissue configuration and Alfonso Martinez Arias for continuous support. 


\clearpage

\appendix
\section*{Supplementary Material}

\noindent MOVIE S1: Time-lapse movie of a wild-type embryo carrying the ArmadilloYFP and sGMCA transgenes during DC; time interval between frames is 10 seconds.

\noindent MOVIE S2: Simulation of AS dynamics from early phase ($t = -50$ min) to fast phase ($t = 80$ min).

\end{document}